\documentclass[12pt]{article}
\usepackage{graphicx}
\begin{document}
\rightline{NKU-09-SF1}
\bigskip
\begin{center}
{\Large\bf    Spinning   Dilaton Black Holes in 2 +1 Dimensions: Quasi-normal Modes  and the Area Spectrum}

\end{center}
\hspace{0.4cm}
\begin{center}
Sharmanthie Fernando \footnote{fernando@nku.edu}\\
{\small\it Department of Physics \& Geology}\\
{\small\it Northern Kentucky University}\\
{\small\it Highland Heights}\\
{\small\it Kentucky 41099}\\
{\small\it U.S.A.}\\

\end{center}

\begin{center}
{\bf Abstract}
\end{center}

\hspace{0.7cm} 

We have studied the   perturbation of a spinning dilaton black hole in 2 +1 dimensions by a massless scalar field. The wave equations of a massless  scalar field is shown to be exactly solvable in terms of hypergeometric functions. The quasinormal frequencies are computed  for slowly spinning black holes. The stability of the black hole is discussed. The asymptotic form of  the quasinormal frequencies are evaluated. The area spectrum of the quantum black holes are evaluated by using the asymptotic quasi-normal frequencies and is shown to be equally-spaced.


\section{Introduction}

The low-energy string effective actions contain  many scalar fields including the dilaton, axion and modulie fields which couples to gravity in a non-trivial manner. Hence the black holes in string theory have modified properties compared to its counterparts in the Einstein's gravity alone. There are many works on black holes in string theory  in literature \cite{ortin}. In this paper we focus on a spinning black hole in gravity coupled to a dilaton with a cosmological constant in 2 +1 dimensions. The action corresponding to this black hole is the low-energy string effective action in 2+1 dimensions.

Low dimensional gravity provides a simpler setting to investigate properties of black holes in a complex theory such as Einstein-dilaton gravity. The well known BTZ black hole in 2 + 1 dimensions \cite{banados} which has attracted much interest has provided insights into many aspects of black hole physics. The first dilaton black hole in 2 + 1 dimensions was derived by Chan and Mann in\cite{chan1}. It was static and was electrically charged. A class of spinning dilaton black holes in 2 +1 dimensions were derived by  Chan and Mann in\cite{chan2}. One of such black holes corresponds to the  low-energy string action  which will be the  focus of the current paper. Modifications of the BTZ black hole by a dilaton field was also considered \cite{chan3}. In an interesting paper, Chen generated new class of  dilaton solutions applying T-duality to known solutions in 2+1 dimensions \cite{chen}. Rotating dilaton solutions were also derived by compactification   of 4D cylindrical solutions by Fernando \cite{fer1}. 

Quasi normal  modes (QNM) arises when a black hole is perturbed by an external field. These  are damped modes with complex frequencies. Such frequencies depend only on the black hole  parameters. The study of QNM's has a long history and there has been extensive work done to compute QNM frequencies  and to analyze them in various black hole backgrounds. A good review is  Kokkotas et. al.\cite{kokko}. QNM's has attracted lot of attention due to  the conjecture relating anti-de-Sitter space (AdS) and conformal field theory(CFT) \cite{aha}. It is conjectured that  the imaginary part of the QNM's which gives the time scale to decay the black hole perturbations  corresponds to the time scale of the conformal field theory (CFT)  on the boundary to reach thermal equilibrium. There are  many works on AdS black holes on this subject \cite{horo1} \cite{car1} \cite{moss} \cite{wang} \cite{sio1} \cite{kon1} \cite{kon2}.

On the other hand, if signals due to QNM's are detected by the gravitational wave detectors, one may be able to identify the charges of black holes and obtain  deeper understanding of the black holes in nature. A  review on QNM's and gravitational wave astronomy written by Ferrari and Gualtieri discuss such possibilities  \cite{ferr}.

QNM's also have been studied in relation to the quantum area spectrum of the black hole horizon. Bekenstein is the first to propose that upon a  suitable quantization process  the area of a black  hole horizon would lead to a discrete spectrum and are evenly spaced \cite{bek1} \cite{bek2}. Hod made an interesting conjecture that the asymptotic QNM frequency and the fundamental area unit in a quantized black hole area indeed are related \cite{hod}. Following Hod's work, Dreyer applied the asymptotic QNM frequency to obtain the  black hole entropy in loop quantum gravity \cite{dre}. Due to these developments, there had been many   works to compute asymptotic QNM frequencies \cite{sio2} and area spectrums in various types of black holes \cite{kun} \cite{set1} \cite{set2} \cite{set3} \cite{set4} \cite{wen} \cite{med} \cite{veg}.

Most of the computations done to evaluate QNM frequencies are numerical due to the nature of the differential equations \cite{kon3}.  However, in 2+1 dimensions, there had been several papers where QNM's are  computed exactly. The well known BTZ black hole \cite{banados} has been studied with exact results \cite{bir1} \cite{bir2} \cite{car2} \cite{abd}. The QNM's of the neutral scalars and charged scalars around the static charged dilaton  black hole in 2 + 1 dimensions were computed exactly  by the current author in \cite{fer2} \cite{fer3}. The Dirac QNM's for the static charged dilaton black hole was computed in \cite{ort3}. In this paper we take a step further by studying QNM's of a massless scalar around a spinning dilaton black hole in 2+1 dimensions. Here, we show that the wave equation can be solved exactly even for a spinning dilaton black hole. We have computed QNM frequencies for  a slowly spinning black hole and given directions to compute QNM frequencies exactly for a general spinning black hole. We will also compute the area spectrum of these black holes.

We have organized the paper as follows: In section 2 an introduction to the geometry of the spinning dilaton black hole is given. The  massless scalar perturbation of the black hole is given  in section 3. The general solution to the wave equation for the  rotating black hole is given in section 4. Solution with boundary conditions is given in section 5. QNM  frequencies  of the black hole is computed and analyzed in detail in section 6. The area spectrum and the quantized entropy of the quantum dilaton black hole is computed in section 7. Finally the conclusion and future directions  are given in section 8.

\section{ Spinning  dilaton black hole}

In this section we will present the geometry and other properties of the spinning dilaton black hole. The Einstein-Maxwell-dilaton action which lead to these black holes considered by Chan and Mann \cite{chan2} is given as follows:
\begin{equation}
S = \int d^3x \sqrt{-g} \left[ R - 4  (\bigtriangledown \phi )^2 -
e^{-4 a \phi} F_{\mu \nu} F^{\mu \nu} + 2 e^{b \phi} \Lambda \right]
\end{equation}
Here, $ \Lambda$ is treated as the cosmological constant, $\phi$ is the dilaton field and $F_{\mu \nu}$   is the Maxwell's field. The black hole considered here is neutral ignoring the Maxwell's field. $ \Lambda > 0 $ corresponds to anti-de-Sitter space and  $\Lambda < 0$ corresponds to the de-Sitter space. A family of spinning black hole solutions characterized by the mass M, angular moment J were presented in \cite{chan2} with the  metric given by,
$$ds^2 = - \left( \frac{ 8 \Lambda r^N}{ (3 N -2)N} + \delta r^{ 1 - \frac{N}{2}}\right) dt^2 + \frac{dr^2}{\left[ \frac{ 8 \Lambda r^N}{ (3 N -2)N} + \left( \delta - \frac{ 2 \Lambda \gamma^2}{ ( 3 N -2)N \delta} \right) r^{ 1 - \frac{N}{2}}\right]}$$
\begin{equation}
- \gamma r^{ 1 - \frac{N}{2}} dt d\theta + \left( r^N - \frac{\gamma^2}{ 4 \delta} r^{ 1 - \frac{N}{2}} \right) d \theta^2
\end{equation}
\begin{equation}
M = \frac{N}{2} \left[\frac{2 \Lambda \gamma^2}{ ( 3 N -2)N \delta} \left( \frac{4}{N} - 3 \right) - \delta \right]; \hspace{0.2cm} J = \frac{ 3 N - 2}{ 4} \gamma
\end{equation}
\begin{equation}
\delta = -\frac { M} { N} - \sqrt{\frac{M^2}{N^2} + \left( \frac{4}{N} - 3 \right) \frac{ 2 \Lambda \gamma^2}{ (3 N -2)N} }
\end{equation}
Here, $\delta$, $\gamma$ are integration constants related to the mass $M$ and the angular momentum $J$ of the black  hole. To avoid closed-time-like co-ordinates, the integration constant $\delta$ must be chosen to be negative. The dilaton field is
\begin{equation}
\phi = k ln( r)
\end{equation}
where,
\begin{equation}
k = \pm \frac{1}{4} \sqrt{ N ( 2 - N)}
\end{equation}
\begin{equation}
bk = N -2
\end{equation}
It was stated in the paper \cite{chan2} that positive mass ( $ M > 0$) black holes exists if $\Lambda >0$ and $ 2 \geq N \geq \frac{2}{3} $. If such conditions are satisfied, the metric in eq.(2) admits an event horizon $r_h$ as,
\begin{equation}
 \left( \frac{ 4}{N} - 3 \right) \frac{ 4 \Lambda }{ ( 3 N - 2) N } r_h^{\frac{3N}{2} -1}  = \frac{ M} {N} \left( \frac{2}{N} -1 \right) + \sqrt{\frac{ 4 M^2}{N^2} + \left( \frac{4}{N} -3 \right) \frac{8 \Lambda \gamma^2}{ (3 N -2)N } } \left( \frac{1}{N} -1 \right)
\end{equation}
Note that this is for $ N \neq 4/3$. If $ N = 4/3$, $r_h$ is given by,
\begin{equation}
3 \Lambda r_h^2 = \frac{ 3 M}{2} - \frac{ \Lambda \gamma^2 }{ 2 M}
\end{equation}

In this paper, we will focus on a special class of black holes with
The values $ b = 4 a = 4$, $ N =1$ and $ k = -1/4$. Such values lead to the low-energy string effective action,
\begin{equation}
S = \int d^3x \sqrt{-g} \left[ R - 4  (\bigtriangledown \phi )^2 -
e^{-4  \phi} F_{\mu \nu} F^{\mu \nu} + 2 e^{ 4 \phi} \Lambda \right]
\end{equation}
The above  action in eq.(9) is related to the low-energy string action in 2+1 dimensions by a conformal transformation,
\begin{equation}
g_{\mu \nu}^{String} = e^{ 4 \phi} g_{\mu \nu}^{Einstein} 
\end{equation}
The spinning black hole solution corresponding to the  action in eq.(9) is given by,
\begin{equation}
ds^2= - f(r)  dt^2 +  \frac{dr^2}{ h(r) } - 4 J r d \theta dt + p(r)^2 d \theta^2
\end{equation}
where,
\begin{equation}
f(r) =  \left(8 \Lambda r^2  - ( M + \sqrt{ M^2 + 32 \Lambda J^2})r\right)
\end{equation}
\begin{equation}
h(r) = \frac{(8 \Lambda r^2  - 2 Mr )}{ 4 r^2 }
\end{equation}
\begin{equation}
p(r)^2 = \left(   r^2 + \frac{( - M + \sqrt{ M^2 + 32 \Lambda J^2} ) }{8 \Lambda} r \right)
\end{equation}
The dilaton field is given by,
\begin{equation}
\phi = -\frac{1}{2} ln(r)
\end{equation}
Note that in deriving the above metric from the one in eq.(2), the value of $\gamma = 4 J$ is substituted to the metric. Also, a simple coordinate transformation has been done to $r$ as $ r \rightarrow r^2 $. Also the constant $\delta$ in the metric of eq.(12) is computed by eq.(4)as,
\begin{equation}
\delta = - M - \sqrt{ M^2 + 32 \Lambda J^2}
\end{equation}
and has been substituted to the metric to eliminate too many constants in the theory.
The black hole has a   horizon given by,
\begin{equation}
r_h =  \frac{M}{4 \Lambda} 
\end{equation}
There is a singularity at $r=0$. The scalar and Kretschmann scalars diverge only at $r =0$. Note that in the presence of a non-trivial dilaton, the space-geometry of the black hole does not behave as either de-Sitter ( $\Lambda < 0$) or anti-de-Sitter ($\Lambda > 0$) \cite{chan2}. An important thermodynamical quantity corresponding to a black hole is the  Hawking temperature $T_H$. It is given by,
\begin{equation}
T_H= \frac{1}{4 \pi} \left( \frac{1}{16 \Lambda^2}  + \frac{J^2}{ M \Lambda( M + \sqrt{ M^2 + 32 \Lambda J^2}) }  \right)^{-1/2}
\end{equation}
The angular velocity at the horizon of the black hole is,
\begin{equation}
\Omega_h  = \frac{ 2 J} { r_h  + r_0 }
\end{equation}
Here,
\begin{equation}
r_0 = \frac{( - M + \sqrt{ M^2 + 32 \Lambda J^2} ) } { 8 \Lambda}
\end{equation}
When the angular moment $J$ approaches zero, the above metric reach the metric of the non-spinning  dilaton black hole discussed in \cite{fer2}. It is interesting to observe that the horizon   of the non-spinning  dilaton black hole and the spinning black hole are the same.

\section{ Massless scalar perturbation of  the spinning dilton  black holes}

In this section, we will develop the equations for a massless scalar field in the background of the spinning  dilaton black hole. The general equation for a massless  scalar field in curved space-time can be written as,
\begin{equation}
\bigtriangledown ^{\mu} \bigtriangledown_{\mu} \Phi =0
 \end{equation}
Using the anzatz,
\begin{equation}
\Phi = e^{- i \omega t} e^{ i m \theta} \frac{\eta(r)} { \sqrt{p(r)}}
\end{equation}
eq.(22)  becomes the  $Schr\ddot{o}dinger$-type equation
\begin{equation}
\frac{ d^2 \eta(r) }{ d r_{*}^2} + \left( \omega^2 - \omega \frac {4mJr}{p(r)^2} - V(r) \right) \eta(r)  =0
\end{equation}
Here, $V(r)$ is given by,
\begin{equation}
V(r) =  -\frac{\sqrt{g(r) h(r) }} {  p(r)^{3/2} } \frac{d}{dr} \left( \sqrt{ g(r) h(r) }   \frac{d}{dr} \left( \frac{1}{\sqrt{p(r)}}\right) \right) 
+ \frac{ m^2 f(r)}{p(r)^2} 
\end{equation}
and $r_{*}$ is the tortoise coordinate computed as,
\begin{equation}
dr_{*} = \frac{p(r) dr}{\sqrt{g(r)h(r)}}
\end{equation}
Here, $g(r)$ is given by,
\begin{equation}
g(r) = f(r) p(r)^2 + 4 r^2 J^2 = r^2 ( 8 \Lambda r^2 - 2 M r )
\end{equation}
When $ J \Rightarrow 0$, the potential $V(r)$ simplifies to,
  \begin{equation}
V(r)_{J=0} =  - \frac{M^2}{4 r}  - \frac{ 2 m^2 M}{ \sqrt{r}} +  8 m^2 \Lambda + 4 \Lambda^2 
\end{equation}
and $r_{*}$ approaches,
\begin{equation}
dr_{*} = \frac{2 r dr}{ 8 \Lambda r^2 - 2 M r} 
\end{equation}
which agrees with the potential and the tortoise coordinate in \cite{fer4} for 
a neutral non-spinning  dilaton black hole. The potentials are plotted in the Fig.1.

\newpage

\begin{center}

\scalebox{0.9} {\includegraphics{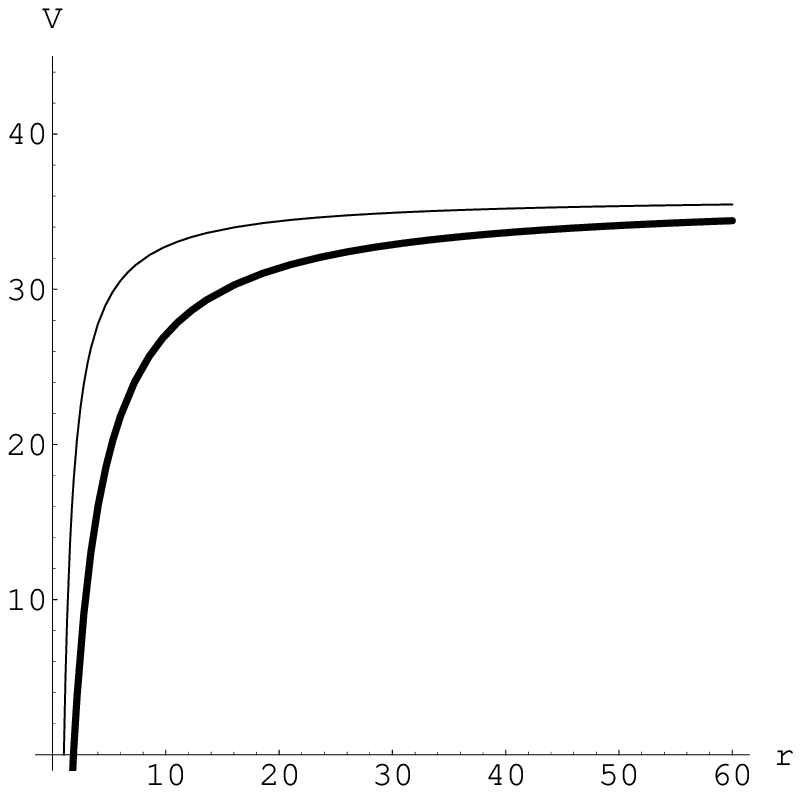}}

\vspace{0.3cm}
\end{center}
{Figure 1. The behavior of the potentials $V(r)$ and $V_{J=0}(r)$ with $r$ for $\Lambda=1$ $M=4$, $J=2$,  and $m = 2$. The dark curve represents $V(r)$ and the light curve represents $V_{J=0}(r)$}

\section { General solution to the  massless scalar wave equation for spinning black holes}

To find exact solutions to the wave equation for the massless scalar, we will revisit the eq.(22) in section 3.
Using the anzatz,
\begin{equation}
\Phi =  e^{- i \omega t} e^{i m \theta} R(r) 
\end{equation}
and re-writing eq.(22) leads to the radial equation,
\begin{equation}
\frac{d}{dr} \left( \sqrt{g(r) h(r)} \frac{dR(r)}{dr} \right) + \left( \frac{ \omega^2 p(r)^2 - m^2 f(r) - 4 \omega m r J }{ \sqrt{g(r) h(r)} }\right) R(r) = 0
\end{equation}
Here,
\begin{equation}
\sqrt{ g(r) h(r) } = 4 \Lambda r^2 - M r  = t(r)
\end{equation}
The function $t(r)$  have two roots given by $r = M/ ( 4 \Lambda) = r_h$ and $r = 0$.

After substituting the functions in eq.(31), the radial equation is simplified to be,
\begin{equation}
\frac{d}{dr} \left( t(r) \frac{dR(r)}{dr} \right) + \left( \frac{ a_1 r}{ t(r)} + \frac{ a_2 r^2}{ t(r) } \right) R(r) = 0
\end{equation}
Here,
$$
a_1 = M \left( -\frac{\omega^2}{ 8 \Lambda} + m^2 \right) + \sqrt{ M^2 + 32 \Lambda J^2}
\left( \frac{\omega^2}{ 8 \Lambda} + m^2 \right)  - 4 \omega m J
$$
\begin{equation}
a_2 =  \omega^2 - 8 \Lambda m^2
\end{equation}
Now, a new variable  is introduced as,
\begin{equation}
z = \frac{ r - r_h}{r }
\end{equation}
Note that in the new coordinate system, $z = 0$ corresponds to $r = r_h$ 
and $z = 1$ corresponds to $ r = \infty$. With the new coordinate,  eq.(33) becomes,
\begin{equation}
z(1-z) \frac{d^2 R}{dz^2} + (1-z) \frac{d R}{dz} + P(z) R =0
\end{equation}
Here,
\begin{equation}
P(z) = \frac{A}{z} + \frac{B}{-1+z} + C
\end{equation}
where,
\begin{equation}
A=  \frac{ ( a_1 + a_2 r_h)}{ 16r_h^2 \Lambda^2}; \hspace{1.0cm} B = \frac{ 8m^2 \Lambda - \omega^2} { 16 \Lambda^2}; \hspace{1.0cm} C = 0; \end{equation}
Now, if   $R(z)$ is redefined as,
\begin{equation}
R(z) = z^{\alpha} (1-z)^{\beta} F(z)
\end{equation}
the radial equation given in eq.(36) becomes,
\begin{equation}
z(1-z) \frac{d^2 F}{dz^2} + \left(1 + 2 \alpha - (1+ 2 \alpha + 2\beta )z \right) \frac{d F}{dz} + \left(\frac{\bar{A}}{z} + \frac{\bar{B}}{-1+z} + \bar{C}     \right) F =0
\end{equation}
where,
$$\bar{A} = A + \alpha^2$$
$$ \bar{B} = B + \beta - \beta^2$$
\begin{equation}
 \bar{C} =  -(\alpha + \beta)^2 
\end{equation}
The eq.(40)  resembles the hypergeometric differential equation  which is of the form \cite{math},
\begin{equation}
z(1-z) \frac{d^2 F}{dz^2} + (c  - (1+a + b )z) \frac{d F}{dz} -ab  F =0
\end{equation}
One can compare  the coefficients of eq.(40) and eq.(42), to obtain the following identities,
\begin{equation}
c = 1+ 2 \alpha
\end{equation}
\begin{equation}
a+b = 2 \alpha + 2 \beta
\end{equation}
\begin{equation}
\bar{A}=A + \alpha^2 =0;  \Rightarrow \alpha = \sqrt{ -A} 
\end{equation}
\begin{equation}
\bar{B} = B + \beta - \beta^2=0; \Rightarrow \beta = \frac{ 1 \pm \sqrt{ 1 + 4 B} }{2}
\end{equation}
\begin{equation}
 ab = -\bar{C} = (\alpha + \beta)^2  
\end{equation}
From eq.(44) and eq.(47),
\begin{equation}
a = b= \alpha + \beta
\end{equation}
By substituting the values of $\alpha$ and $\beta$ from eq.(45) and eq.(46), the values of $a$, $b$, and $c$ can be determined in terms of the parameters in the theory. With such values, the solution to the hypergeometric function $F(z)$ is given by \cite{math},
\begin{equation}
F(a,b,c;z) = \frac{\Gamma(c)} {\Gamma(a) \Gamma(b)} \Sigma \frac{ \Gamma(a+n) \Gamma( b+n)}{ \Gamma(c+n)}  \frac{z^n}{n!}
\end{equation}
with a radius of convergence being the unit circle $|z| =1$. Hence the general solution to the radial part of the charged scalar wave equation is given by,
\begin{equation}
R(z) = z^{\alpha} (1-z)^{\beta} F(a,b,c;z)
\end{equation}
with $a$, $b$, and $c$ given in the above equations. The general solution for the massless wave scalar equation is,
\begin{equation}
\Phi( z, t, \theta) = z^{\alpha} (1-z)^{\beta} F(a,b,c;z) e^{  i m \theta} e^ { - i \omega t}
\end{equation}
Hence, the wave equation for a general value of $J$ can be solved exactly in terms of hypergeometric functions.

\section{Solution to the wave equation for slowly spinning black holes with boundary conditions}

In this section we will obtain solutions to the massless scalar with the boundary condition that the wave is purely ingoing at the horizon. The solutions are analyzed closer to the horizon and at infinity to obtain exact results for the wave function. This is similar to the approach followed in \cite{fer3} but we will describe the details here for the sake of completeness.

In order to facilitate computations, we will restrict ourselves to small values of J; that is we will assume $ J \ll M $. In order to solve for $\alpha$ and $\beta$ in eq.(45) and eq.(46), we will further simplify the expressions for  $a_2$, $\Omega_h$  and $A$ by expanding them  around $J$. By eliminating the $J^2$ terms,  those functions are approximated as,
\begin{equation}
a_2 \approx 2 M m^2 - 4 \omega m J;
\end{equation}
\begin{equation}
\Omega_h \approx \frac{ 2 J}{r_h}
\end{equation}
leading to,
\begin{equation}
A \approx \frac{ \omega ( - 16 J m \Lambda + M \omega )}{ 16 M \Lambda^2} \approx
\frac{ ( \omega - \frac{ 2 J m}{ r_h} )^2}{ 16 \Lambda^2} = \frac{ ( \omega - m \Omega_h)^2}{ 16 \Lambda^2} = \frac{ \hat{\omega}^2}{ 16 \Lambda^2}
\end{equation}
Here, $ \hat{\omega}=\omega - m \Omega $. Now, from eq.(45), $\alpha$ is solved to be,
\begin{equation}
\alpha = \pm \frac { i \hat{ \omega}}{ 4 \Lambda}
\end{equation}
The massless scalar wave equation for non-spinning blackhole was proved to be exactly solvable in the paper by Fernando \cite{fer2}. There, the analogous values for $\alpha$ was given as,
\begin{equation}
\alpha = \pm \frac{ i \omega}{ 4 \Lambda}
\end{equation}
This is indeed the case when $ J \rightarrow 0$ in the value obtained for the spinning black hole here.

From eq.(46) and $B$ in eq.(38),
\begin{equation}
\beta = \frac{1 +  i \sqrt{ ( \frac{ \omega^2 - 8 m^2 \Lambda}{4 \Lambda^2} - 1 }) }{2}
\end{equation}

\subsection{ Solution at the near-horizon region}

First, the solution of the wave equation closer to the horizon is analyzed. For the spinning  black hole, as the radial coordinate $r$ approaches the horizon, $z$ approaches $0$. In the neighborhood of $z=0$, the hypergeometric function has two linearly independent solutions given by \cite{math}
\begin{equation}
F(a,b;c;z) \hspace{1.0cm} and \hspace{1.0cm} z^{(1-c)} F(a-c+1,b-c+1;2-c;z)
\end{equation}
Substituting the  values of $a,b,c$ in terms of $\alpha$ and $\beta$ obtained in eq.(54) and (57), the general solution for $R(z)$ can be written as,
$$
R(z) = C_1 z^{\alpha} (1-z)^{\beta} F(\alpha + \beta , \alpha+\beta , 1+ 2 \alpha, z) 
$$
\begin{equation}
+ C_2 z^{-\alpha}(1-z)^{\beta} F( -\alpha + \beta ,-\alpha+\beta ,1-2 \alpha, z)
\end{equation}
Here, $C_1$ and $C_2$ are constants to be determined. We want to  point out that the above equation is symmetric for $ \alpha \leftrightarrow - \alpha$. Note that in eq.(55), $\alpha$ could have both $\pm$ signs. Hence due to the above symmetry, we will choose the ``+" sign for $\alpha$  for the rest of the paper.

Since  closer to the horizon $z \rightarrow 0$, the above solution in eq.(59) approaches,
\begin{equation}
R(z  \rightarrow 0) = C_1 z^{\alpha}  + C_2 z^{-\alpha}
\end{equation}
Closer to the horizon, $r \rightarrow r_h$. Hence, $z$ can be approximated with
\begin{equation}
z \approx \frac{ r - r_h}{r_h}
\end{equation}
The ``tortoise'' coordinate for the slowly rotating black hole near the horizon can be approximated as,
\begin{equation}
dr_{*} \approx \frac{dr}{ 4 \Lambda ( r - r_h)}
\end{equation}
leading to,
\begin{equation}
r_* \approx  \frac{1}{4 \Lambda } ln( r  - r_h)
\end{equation}
Hence,
\begin{equation}
r - r_h = e^{ 4 \Lambda  r_* }
\end{equation}
leading to,
\begin{equation}
z \approx  \frac{ 1 }{ r_h} e^{ 4 \Lambda  r_* }
\end{equation}
Hence eq.(60) can be re-written in terms of $r_*$ as,
\begin{equation}
R(r \rightarrow r_+) =  C_1 \left(\frac{ 1} { r_h} \right)^{\alpha} e ^{ i  \hat{\omega} r_*}  + C_2 \left(\frac{1}{ r_h} \right)^{ - \alpha} e^{ -i  \hat{\omega} r_*}
\end{equation}
To obtain the above expression, $\alpha$  is substituted from eq.(55) and
\begin{equation}
\hat{\omega} = \omega - \frac{ 2 m J } { r_h}
\end{equation}
The first and the second term in eq.(66) corresponds to the outgoing and the ingoing wave respectively. Now, by imposing  the condition  that  the wave is purely ingoing at the horizon, one can   pick $C_1 = 0$ and $C_2 \neq 0$. Therefore the solution closer to the horizon is,
\begin{equation}
R(z \rightarrow 0 ) =  C_2 z^{-\alpha} (1-z)^{\beta} F(-\alpha + \beta , -\alpha+\beta , 1 - 2 \alpha, z) 
\end{equation}

\subsection{Solution at asymptotic region}

In this section  the wave equation in eq.(33) is analyzed when $r \rightarrow \infty$. For large $r$, the function $t(r) \rightarrow 4 \Lambda r^2$. When $t(r)$ is replaced with this approximated function in the wave equation given by eq.(33), it simplifies to,

\begin{equation}
\frac{d}{dr} \left( 4 \Lambda r^2 \frac{dR(r)}{dr} \right) + 2r^2 \left( \frac{\omega^2}{ 8 \Lambda r^2}   - \frac{m^2}{r^2}  \right)  R(r) = 0
\end{equation}
Here, we have neglected the other terms since $r$ is large. Hence finally, the wave equation at large $r$ can be  expanded to be,
\begin{equation}
r^2 R'' + 2 r R' + \mu R =0
\end{equation}
where, 
\begin{equation}
\mu = \frac{\omega^2} { 16 \Lambda^2} - \frac{m^2}{2 \Lambda}
\end{equation}
One can observe that $ \mu = -B$ from eq.(38). Also eq.(70) is the well known Euler equation with the solution,
\begin{equation}
R(r) = D_1 \left( \frac{r_h}{r} \right)^{\nu_1} + D_2 \left( \frac{r_h }{r} \right)^{\nu_2}
\end{equation}
with,
\begin{equation}
\nu_1=   \frac{ 1 + \sqrt{ 1 - 4 \mu} }{2} =  \beta; \hspace{1.0 cm}
\nu_2 =  \frac{ 1 - \sqrt{ 1 - 4 \mu} }{2} = (1- \beta)
\end{equation}
The expression for $\beta$ is given in eq.(46). Note that the form in eq.(72) is chosen to facilitate to compare it with the matching solutions in section 5.3.

\subsection{Matching the solutions at the near horizon and the asymptotic region}

In this section we  match the asymptotic solution given in eq.(72) to the large $r$ limit (or the $z \rightarrow 1$ ) of the near-horizon solution given in 
eq.(68) to obtain an exact expression for $D_1$ and $D_2$.  To obtain the $z \rightarrow 1$ behavior of eq.(68), one can perform a transformation on hypergeometric function given as follows \cite{math}
$$
F(a,b,c,z) = \frac{ \Gamma(c) \Gamma(c-a-b)}{\Gamma(c-a) \Gamma(c-b)} F(a,b;a+b-c+1;1-z) 
$$
\begin{equation}
 +(1-z)^{c-a-b}\frac{ \Gamma(c) \Gamma(a+b-c)}{\Gamma(a) \Gamma(b)} F(c-a,c-b;c-a-b+1;1-z)
\end{equation}
Applying this transformation to eq.(68) and substituting for the values of $a,b,c$, one can obtain the solution to the wave equation in the asymptotic region as follows;
$$
R(z) = C_2 z^{-\alpha} (1-z)^{\beta} \frac{ \Gamma(1 - 2 \alpha) \Gamma(1 - 2 \beta)}{\Gamma(1 - \alpha - \beta )^2} F( -\alpha + \beta , -\alpha + \beta; 2 \beta ;1-z) $$
\begin{equation}
+ C_2  z^{-\alpha} (1-z)^{1 - \beta} \frac{ \Gamma( 1 - 2 \alpha ) \Gamma( -1 + 2 \beta )}{\Gamma( -\alpha + \beta )^2} F( 1 - \alpha - \beta , 1 - \alpha - \beta  ; 2 - 2 \beta;1-z)
\end{equation}
Now we can take the limit of  $R(z)$ as $ z \rightarrow 1$ ( or $r \rightarrow \infty$) which will lead to,
$$
R(z \rightarrow 1) = C_2  (1-z)^{\beta} \frac{ \Gamma(1 - 2 \alpha) \Gamma(1 - 2 \beta)}{\Gamma(1 - \alpha - \beta )^2} 
$$
\begin{equation}
+ C_2 (1-z)^{1 - \beta} \frac{ \Gamma( 1 - 2 \alpha ) \Gamma( -1 + 2 \beta )}{\Gamma( -\alpha + \beta )^2 } 
\end{equation}
Note that we have replaced $F(a,b,c,1- z)$ and $z^{\alpha}$ with 1  when $z$ approaches 1.
Since,
\begin{equation}
1 - z = \frac{r_h}{r},
\end{equation}
by replacing $1 - z$ with the above expression in eq.(76), $R(r)$ for large $r$ can be written as,
$$
R(r \rightarrow \infty) = C_2  \left(\frac{r_h}{r }\right)^{\beta} \frac{ \Gamma(1 - 2 \alpha) \Gamma(1 - 2 \beta)}{\Gamma(1 - \alpha - \beta )^2}
$$
\begin{equation} 
+ C_2  \left(\frac{r_h}{r }\right)^{1 - \beta} \frac{ \Gamma( 1 - 2 \alpha ) \Gamma( -1 + 2 \beta )}{\Gamma( -\alpha + \beta )^2 } 
\end{equation}
By comparing eq.(72) and eq.(78), the coefficients $D_1$ and $D_2$ can be written as,
\begin{equation}
D_1 = C_2 \frac{ \Gamma(1 - 2 \alpha) \Gamma(1 - 2 \beta)}{\Gamma(1 - \alpha - \beta )^2 } 
\end{equation}
\begin{equation}
D_2 = C_2  \frac{ \Gamma( 1 - 2 \alpha ) \Gamma( -1 + 2 \beta )}{\Gamma( -\alpha + \beta )^2} 
\end{equation}
To determine which part of the solution in eq.(72) corresponds to the ``ingoing'' and ``outgoing'' respectively, we will first find the tortoise coordinate $r_{*}$ in terms of $r$ at large r. Note that for large $r$, $t(r) \rightarrow 4 \Lambda r^2 $. Hence  the equation relating the tortoise coordinate $r_*$ and $r$ in eq.(62) simplifies to,
\begin{equation}
dr_{*} = \frac{ dr}{ 4 \Lambda r}
\end{equation}
The above can be integrated to obtain,
\begin{equation}
r_{*} \approx  \frac{1}{4 \Lambda} ln( \frac{ r}{r_h} )
\end{equation}
Hence,
\begin{equation}
r \approx r_h e^{ 4 \Lambda r_*}
\end{equation}
Substituting $r$ from eq.(83) and $\beta$ from eq.(57) into the eq.(72), $R(r \rightarrow \infty)$ is rewritten as,
$$
R(r  \rightarrow \infty ) \rightarrow D_1 r_h^{\beta}  e ^{ -i  \omega r_* \sqrt{1 - \frac{ 4 \Lambda^2}{ \omega^2} ( \frac{2 m^2}{\Lambda} +1) } - 2 \Lambda r_*} $$
\begin{equation}
+ D_2  r_h^{(1 - \beta)} e ^{ i  \omega r_* \sqrt{1 - \frac{ 4 \Lambda^2}{ \omega^2} ( \frac{2 m^2}{\Lambda} +1) } - 2 \Lambda r_*} 
\end{equation}
From the above it is clear that the first term and the second term represents the ingoing and outgoing waves respectively.

\section{Quasinormal modes of the spinning dilaton black hole}

When computing the quasi normal modes of a classical perturbation of black hole space-times,  two boundary conditions are imposed. First is  the condition that the waves are purely ingoing at the horizon which is already done in the computations in section 5. In addition, one has to impose boundary conditions on the solutions at the asymptotic region as well. In asymptotically flat space-times, the second boundary condition is the solution to be purely outgoing  at spatial infinity. For non-asymptotically flat space times, there are two possible boundary conditions to impose at sufficiently large distances from the black hole horizon: one, is the field to vanish at large distances and the other is for the flux of the field to vanish at far from the horizon. The first is chosen here. This is the condition imposed in reference \cite{fer2}. Another example in 2+1 dimensions where the vanishing of the field at large distance is imposed is given in reference \cite{bir1} where QNM's of scalar perturbations of BTZ black holes were computed exactly.

Let's consider the field $R(r)$ at large distances given by eq.(84). Clearly the second term vanishes when $ r \rightarrow \infty$. This  also can be seen from eq.(76) where the second term vanishes for $ z \rightarrow 1$. Since $C_2$ is not zero, the first term vanish only at the poles of the Gamma function $\Gamma(1 - \alpha - \beta )$. Note that the Gamma function $\Gamma(x)$ has poles at $ x = - n$ for $ n = 0,1,2..$.Hence to obtain QNM's, the following relations has to hold.
\begin{equation}
1- \alpha - \beta = - n
\end{equation}
which leads to the equation for $\beta$ given by,
\begin{equation}
\beta = ( 1 + n) - \frac{ i (\omega - \frac{ 2 mJ}{rh}) } { 4 \Lambda }
\end{equation}
By combining the  above equation with the eq.(46) given by,
\begin{equation}
\frac{m^2}{  2 \Lambda} - \frac{ \omega^2}{ 16 \Lambda^2} + \beta - \beta^2 = 0
\end{equation}
$\omega = \omega_R + i \omega_I$ can be found where,
\begin{equation}
\omega_R =  \frac{ 2 J m}{\Lambda r_h (2n + 1)^2} \left(  \Lambda ( 1 + 2 n + 2 n^2) + m^2 \right)
\end{equation}

\begin{equation}
\omega_I =  \frac{-2 }{ 2n +1} \left( 2 \Lambda n (1+n) - m^2 \right)
\end{equation}
Due to the minus sign in front of $\omega_I$, these oscillations will be damped leading to stable perturbations for $2 \Lambda n (1+n) > m^2$. However, for $2 \Lambda n (1+n) < m^2$, the oscillations would lead to unstable modes. This was pointed out in \cite{ort3}. In fact, for non-rotating black holes, the $\omega_I$ is same as for the slowly rotating black hole. However, $\omega_R$ was zero for the non-rotating black hole where as here it is non zero and depend on $J$.

The asymptotic values of the $\omega$ is computed as follows;

\begin{equation}
\omega_{R} ( n \rightarrow \infty ) = \frac{  J m}{ r_h}
\end{equation}

\begin{equation}
\omega_{I} ( n \rightarrow \infty ) = - 2 n \Lambda
\end{equation}

\begin{center}

\scalebox{0.9} {\includegraphics{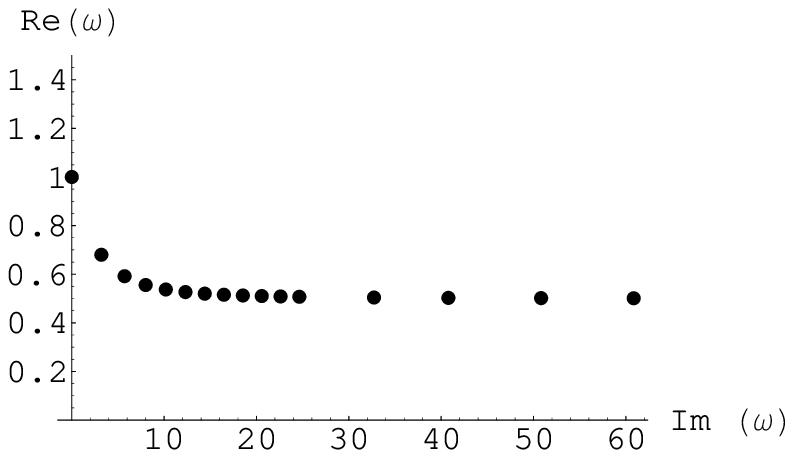}}

\vspace{0.3cm}
\end{center}
{Figure 2. $\omega_{real}$ against $\omega_{imaginary}$ for $\Lambda=1$ $M=32$, $J=2$,  and $m = 2$ }\\

In the above figure, for large $n$, $\omega_R$ approaches a constant value and $\omega_I$ becomes linear in $n$.

\section{ Area spectrum of the spinning dilaton black hole}

The relation of quasi normal mode frequencies and the quantization of the
area spectrum of a black hole is been studied extensively. Following the methods of previous work, we will compute  the area spectrum and the entropy of the dilaton black hole considered in this paper.

Bekenstien is the first to conjecture that in a quantum theory, the black hole area would be discrete and equally spaced \cite{bek1} \cite{bek2}. Hod made an interesting conjecture that the asymptotic QNM frequency and the fundamental area unit in a quantized black hole area indeed are related \cite{hod}. Following Hod's work, Dreyer applied the asymptotic QNM frequency to obtain the  black hole entropy in loop quantum gravity \cite{dre}. Birmingham et.al. showed that Hod's ideas could be translated into interesting statements about the dual CFT in \cite{carlip}. Due to these developments, there had been many   works to compute asymptotic QNM frequencies and area spectrums in various types of black holes.

In an interesting paper, Kunstatter stated that \cite{kun},
\begin{equation}
I = \int \frac{ dE} {  \omega(E) }
\end{equation}
is an adiabatic invariant for a system with energy $E$ and vibrational
frequency $\omega(E)$. Furthermore it was stated that via the Bohr-Sommerfeld quantization, the quantity $I$ will have an equally spaced spectrum in the large $n$ limit as,
\begin{equation}
I \approx n \hbar
\end{equation}
For black holes, the vibrational frequency was the QNM frequency and
$dE$ was obtained using the first Law of black hole thermodynamics. Kunstatter, applied this approach to compute the black hole entropy spectrum for d-dimensional Schwarzschild black holes in \cite{kun}. The $  \omega(E)$ was substituted with the real part of the quasi-normal frequency at the large $n$ limit.

There has been several interesting work following this approach to quantize the area of  various black holes in the literature. For example, the area spectrum of the non-spinning BTZ black hole \cite{set1}, the extremal Reissner-Nordstrom black hole\cite{set2}, Kerr black hole \cite{set3}, extremal Schwarzschild-de-Sitter black hole \cite{set4} were computed in the respective papers. In all these works, the $ \omega(E) $ in the expression in eq.(92) was chosen to be the real part $\omega_R$ of the quasi-normal frequency at large $n$ limit. However, in a recent paper, Maggiore \cite{mag} has argued that, in the large $n$ limit ( or high damping limit), the physically relevant $\omega$ would be,
\begin{equation}
\omega_p = \sqrt{ |\omega_R|^2 + |\omega_I|^2 }
\end{equation}
However, it was also stated that the expression $\omega$ in the eq.(92) should be the transition between (adjacent) quasi-normal frequencies for large n, such that,
\begin{equation}
\omega_t = \lim_{n \rightarrow \infty}  \omega_p( n + 1 ) - \omega_p(n)
\end{equation}
Applying the above idea, area spectrum of various black holes have been re-calculated leading to interesting results. Four papers along these lines are
\cite{ wen} \cite{med} \cite{veg} \cite{li}.

In this paper, we like to calculate the area spectrum of the spinning dialton black holes by extending Kunstatter's and Maggiore's approaches. Note that according to the eq.(89), the oscillations will be damped leading to stable modes for $ 2 \Lambda n( n + 1) > m^2$. Even though the low-lying modes are unstable with $ 2 \Lambda n ( n + 1)  < m^2$,  the highly exited black holes with the limit $ n \rightarrow \infty$,    are stable with
$ \omega_I \rightarrow - 2 n \Lambda$. Since in the approach by Kunstatter and Maggiore, the area is quantized only using the asymptotic QNM frequencies, we will follow their formalism to compute the area spectrum for the black hole considered in this paper. Furthermore, to justify quantizing the area of the black hole in spite of existing unstable modes, we may recall that a full analysis of the QNM's of the spinning black holes in this paper has to be studied for arbitrary values of $J$ to conclude about the stability of the black hole. Since we have only obtained QNM's for a slowly spinning black hole, the computations in the  next section may be considered as an initial step to understand the quantum properties of a possibly stable black hole.

\subsection{ Non-spinning black hole area spectrum}

Kunstatter  incorporated the First Law of thermodynamics in describing the adiabatic invariant $I$ for the Schwarzschild black hole in \cite{kun}. Following the approach in \cite{kun} $dE = dM$ for the non-spinning dilaton black hole. Here,  $\omega$ in the expression in eq.(92) is replaced with $\omega_t$ following Maggiore \cite{mag}. Hence the adiabatic expression for the non-spinning black hole is,
\begin{equation}
I = \int \frac{ d M}{  \omega_t}
\end{equation}
It was shown that the  QNM for the non-spinning dilaton black hole was pure imaginary in \cite{fer2}. Hence, according to Maggiore, the corresponding frequency would be $\omega_p = |\omega_I|$. In the large $n$ limit, it is given by
$2 n \Lambda$. Hence  the transition frequency, $\omega_t$ is,
\begin{equation}
\omega_t = |(\omega_I)_n| - |(\omega_I)_{n -1}| = 2 \Lambda
\end{equation}
Hence the adiabatic invariant for the  non-spinning dilaton black hole is,
\begin{equation}
 I = \int \frac{d M}{2 \Lambda} = \frac{M}{2 \Lambda} 
\end{equation}
From the Bohr-Sommerfeld quantization,
\begin{equation}
I = \frac{M}{2 \Lambda} = n \hbar
\end{equation}
Since the horizon of the non-spinning black hole is $r_h = \frac{M}{4 \Lambda}$, the area spectrum is derived as,
\begin{equation}
A_n = 2 \pi r_h = \pi n \hbar
\end{equation}
It is obvious that the area spectrum is equally spaced. The corresponding entropy spectrum is given by,
\begin{equation}
S_n = \frac{ A_n}{4} = \frac{\pi}{4} n \hbar
\end{equation}
The entropy too is equidistant. 

\subsection{ The Spinning dilaton black hole area spectrum}

Extending Kunstatter's argument for the Schwarzschild black hole, Setare and Vagenas computed the area spectrum for the Kerr black hole \cite{set3}. There, the argument was made that the correct expression for the adiabatic invariant would be
\begin{equation}
I = \int \frac{ d M - \Omega_h d J}{  \omega_t}
\end{equation}
Here again the first law of  thermodynamics was used and the modification was necessary due to the spinning of the black hole. From eq.(90) and eq.(91), in the large $n$ limit, $ \omega_I$ $\gg$ $\omega_R$. Hence, the transition frequency $\omega_t$ is given by 
\begin{equation}
\omega_t = |(\omega_I)_n| - |(\omega_I)_{n -1}| = 2 \Lambda
\end{equation}
Hence the adiabatic invariant for the  slowly spinning dilaton black hole is,
\begin{equation}
 I = \int \frac{d M}{2 \Lambda} - \frac{ 4  J dJ}{ M} = \frac{ M}{2 \Lambda} - \frac{ 2 J^2}{M} 
\end{equation}
Since we have neglected $J^2$ terms for slowly spinning black holes in this paper, one can approximate $I \approx \frac{M}{ 2 \Lambda}$. Hence from the Bohr-Sommerfeld quantization,
\begin{equation}
I = \frac{M}{2 \Lambda}  = n \hbar
\end{equation}
Since the horizon of the spinning black hole is $r_h = \frac{M}{4 \Lambda}$, the area spectrum is derived as,
\begin{equation}
A_n = 2 \pi r_h = \pi n \hbar
\end{equation}
It is obvious that the area spectrum is equally spaced. The corresponding entropy spectrum is given by,
\begin{equation}
S_n = \frac{ A_n}{4} = \frac{\pi}{4} n \hbar
\end{equation}
The entropy too is equidistant similar to the non-spinning black hole. It is interesting to note that both the  non-spinning and spinning black hole
have the same spectrum.

\section{Conclusion}

We have studied {\it analytically} the perturbation of the spinning dilaton black hole in 2+1 by a massless scalar. The wave equation is solved exactly in terms of   hypergeometric functions for a spinning dilaton black hole. To the authors knowledge, this is the only spinning dilaton black hole with exact solutions to the wave equation. The QNM frequencies  are computed  for slowly spinning  black holes. The QNM's have both a real and an imaginary component. The black hole is stable  only for a selected range of values of $n$ and $m$. In a previous paper by the same author \cite{fer1}, the non-spinning dilaton black hole perturbations by a massless scalar field were studied. There, the wave equations were solved exactly and the QNM's were shown to pure imaginary. We have computed the area spectrum of the spinning as well as the non-spinning dilaton black hole which is shown to be equally spaced.

There are several avenues to proceed  from here. Since we have only studied the {\it slowly spinning} black hole, it would be interesting to compute the QNM frequencies for a dilaton black hole with any value of $J$. Such an approach would give a complete solution  to the question of stability of a  spinning black hole  discussed here. If one proceed to compute QNM's for a general spinning black hole following the approach in section 5 and section 6, one will end up with the eq. (85). By substituting the value of   $A$ in eq.(38) without approximations to compute $\alpha = \sqrt{-A}$ and combining with eq.(87) will lead to a polynomial in terms of  $\omega$. Such a polynomial   can be solved numerically.

Another avenue to proceed is to study maasive scalar field around the black hole in this paper. Even though Kerr black hole is stable under massless scalar field, instability has been shown to occur for massive scalar field around it due to superradiance effect \cite{dolan}. It would be interesting to understand if such phenomenon occur in the spinning dilaton black hole.

Also, it would be interesting how the Dirac spinor fields decay around such black holes. One would hope that exact solutions to the Dirac field equations may exist similar to the scalar field. The QNM's of the Dirac field for the non-spinning black hole was computed exactly in \cite{ort3}.

It would be interesting to compute the greybody factors and particle emission rates for the massless scalars for this black hole. The greybody factors of the massless scalar was  studied for the non-spinning black hole in  \cite{fer4}. Since the wave equation is been already solved, it should be a welcome step towards  understanding the Hawking radiation from these black holes. 

In string theory, different space-time geometries may be related to each other by duality transformations. The non-spinning dilaton black hole and the spinning dilaton black hole are related by a T-duality as described by Chen \cite{chen}. Furthermore, the static charged dilaton black hole was generated  by applying the duality transformations described by Horowitz \cite{horo2} to the non-spinning dilaton black hole in \cite{fer2}. In \cite{fer2}, the QNM's were computed for both static charged and neutral black holes. Since QNM frequencies for all three families have been computed, it would be interesting to study what role QNM's play in understanding duality. What remains is to compute QNM's of the spinning charged dilaton black holes in 2 + 1 dimensions. Such solutions were presented by Chen \cite{chen}.

\end{document}